\documentclass[pra,onecolumn,superscriptaddress,aps]{revtex4}
\usepackage{amsfonts}
\usepackage{amssymb,latexsym,amsmath}
\usepackage{graphicx}
\usepackage{times}
\usepackage[caption=false]{subfig}
\usepackage[usenames]{color}

\begin{document}

\title{Asymptotic Expansions and Solitons of  
the Camassa-Holm -- Nonlinear Schr{\"o}dinger Equation
}


\author{I. K. Mylonas}
\affiliation{Department of Mechanical Engineering, Faculty of Engineering Aristotle
	University of Thessaloniki, 54124 Thessaloniki, Greece}

\affiliation{Department of Mathematics and Statistics, University of
	Massachusetts, Amherst MA 01003-4515, USA}

\author{C. B. Ward}
\affiliation{Department of Mathematics and Statistics, University of
	Massachusetts, Amherst MA 01003-4515, USA}

\author{P. G. Kevrekidis}
\affiliation{Department of Mathematics and Statistics, University of
	Massachusetts, Amherst MA 01003-4515, USA}

\author{V. M Rothos}
\affiliation{Department of Mechanical Engineering, Faculty of Engineering Aristotle
	University of Thessaloniki, 54124 Thessaloniki, Greece}

\author{D. J. Frantzeskakis}
\affiliation{Department of Physics, University of Athens,
	Panepistimiopolis, Zografos, Athens 15784, Greece}


\begin{abstract}

We study a deformation of the defocusing nonlinear Schr\"{o}dinger (NLS) equation, 
the defocusing Camassa-Holm NLS, hereafter referred to as CH-NLS equation. 
We use asymptotic multiscale expansion methods to reduce this model to a Boussinesq-like 
equation, which is then subsequently approximated by 
two Korteweg-de Vries (KdV) equations for left- and right-traveling waves. We use 
the soliton solution of the KdV equation to construct approximate solutions of the CH-NLS system. 
It is shown that these solutions may have the form of either dark or antidark solitons, 
namely dips or humps on top of a stable continuous-wave background. 
We also use numerical simulations to investigate the validity of the asymptotic solutions, 
study their evolution, and their head-on collisions. It is shown that small-amplitude 
dark and antidark solitons undergo quasi-elastic collisions. 

\end{abstract}

\maketitle
        
\section{Introduction}

The Camassa-Holm (CH) equation~\cite{ch} is a completely integrable 
nonlinear dispersive partial differential equation (PDE), 
which has attracted much interest due to its soliton solutions, 
its applications to shallow water waves~\cite{l14,l27}, and its interesting mathematical 
features, such as the long-time asymptotics~\cite{l6}, and inverse scattering~\cite{gerdj}. 
It is worthwhile to note that the applications of the model are not exhausted in the theme 
of water waves, but rather have extended more also to areas such as 
acoustic scattering~\cite{l5}, as well as axial deformation waves arising 
in hyperelastic rods~\cite{l16,l17}, among others. 

On the other hand, the nonlinear Schr{\"o}dinger (NLS) equation 
\cite{ablo91,ablowitz1,ablo2011,sulem,chap01:bourgain}
is a principal completely integrable nonlinear dispersive PDE, that has 
been fundamental towards shaping our understanding in a diverse array of systems. These range
from atomic physics~\cite{rcg:BEC_BOOK,rcg:BEC_book2} and nonlinear
optics~\cite{ablo2011,hasegawa,kivshar}, to plasmas~\cite{infeld}, 
deep water waves \cite{ablo2011,johnson74}, rogue waves~\cite{onofrio}, and so on. 

In an intriguing recent development, the work of~\cite{Arnaudon}
proposed a novel deformation of the NLS equation, as a member of
a series of deformations of integrable equations. The corresponding 
deformation of the famous Korteweg-de Vries (KdV) equation
produced the CH equation, hence it is of particular interest
to study the relevant deformation of the NLS equation, i.e., the
CH-NLS equation. In Ref.~\cite{Arnaudon2} the focusing CH-NLS system was studied 
both analytically and numerically, and its bright soliton solutions and their dynamics 
and interactions were explored; note, however, that in the work~\cite{Arnaudon2} no 
definitive conclusion about the complete integrability of the CH-NLS equation was reported.

Our aim in the present work is to explore the defocusing variant of
the CH-NLS equation, and study its dark soliton solutions. 
Given the lack of information regarding the integrability of this model and 
the existence of exact analytical dark soliton solutions, we resort to asymptotic 
multiscale expansion techniques \cite{jk}. Such techniques have been used to 
demonstrate that several completely integrable systems can be reduced to other integrable
models \cite{zakharov86}. Importantly, these techniques have also 
been extended to the case of nonintegrable systems: in such cases, solutions 
of the reduced models can be used to construct approximate solutions of the original models.  
For instance, variants of the defocusing NLS equation have been 
reduced to the KdV equation, which allowed for the description of shallow
dark solitons of the former in terms of KdV solitons (see, e.g., the reviews \cite{ybld,djfr} 
and references therein). This approach also 
allowed for the prediction of the existence of a structure known as {\it antidark soliton}, 
namely a dark soliton with reverse-sign amplitude, having the form of a hump (instead
of a dip) on top of the continuous-wave (cw) background density \cite{kivshar90,yuriol,kiv91,kal,m1}.
Note that relevant studies employing multiscale expansion methods and 
predicting the occurrence of antidark solitons were recently extended to settings 
involving nonlocal nonlinearities \cite{Hor1,Hor2,Hor3}.

A brief description of our findings, as well as the outline of 
the presentation, is as follows. 
In Section~2, we present the model and briefly 
revisit its modulational stability analysis~\cite{Arnaudon2}.
In Section~3, we use multiscale expansion methods to
derive asymptotic reductions of the defocusing CH-NLS equation. 
At an intermediate stage of the asymptotic analysis, we obtain 
a Boussinesq-type equation; next, we consider the far-field of
the Boussinesq equation, and derive two KdV equations for
right- and left-going waves. We then employ the KdV soliton to 
construct approximate soliton solutions of the original CH-NLS equation; 
we show that these soliton solutions are either of the dark or
of the antidark type, 
depending on the sign of a characteristic parameter. 
In Section~3, we present our numerical computations and results of
direct numerical simulations. 
We calculate the error between the asymptotic solutions and the 
numerical solutions, and then we focus on the numerical study of the 
dynamics and head-on collisions of the dark and antidark solitons. 
It is found that such collisions are quasi-elastic for solitons of
sufficiently 
small amplitude. We also show that collisions between different soliton
types, i.e., 
between dark and antidark solitons are also possible and discuss the
special conditions under which these can be realized. 
Finally, in Section~4, we summarize our findings and present our conclusions,
as well as suggest a number of directions for future study.

\section{Model and continuous-wave solution}

As indicated above, our aim is to study the CH-NLS equation. This model was derived in Refs.~\cite{Arnaudon,Arnaudon2} when developing a theory of a deformation of hierarchies 
of integrable systems. In fact, the CH-NLS equation is the deformation of the
NLS equation, in the same sense as the CH equation is the deformation of the KdV equation.
In its standard dimensionless form, the CH-NLS equation reads:
\begin{equation}\label{1}
	i m_{t} + u_{xx} +2\sigma m (|u|^2-a^2 |u_{x}|^2)=0, 
	\qquad m=u-a^2 u_{xx}, 
	\end{equation}
where subscripts denote partial derivatives, $u(x,t)$ is a complex field, $\sigma= \pm 1$ pertains,  respectively, to focusing or defocusing nonlinearity, while $a$ is a constant 
arising within the Helmholtz operator. Obviously, for $a=0$ the above model reduces 
to the standard, completely integrable NLS equation. 
In terms of the complex field $u$, the CH-NLS equation can be expressed as:
	\begin{equation}\label{2}
	i u_{t}+u_{xx}+2\sigma u|u|^{2}-i a^2 u_{xxt}-2\sigma a^{2}u|u_{x}|^2 -2\sigma a^{2} u_{xx}|u|^{2} +2\sigma a^{4}u_{xx}|u_{x}|^{2}=0. \quad
	\end{equation}
	
To start our analysis, we use the Madelung transformation
$u(x,t)=u_{0} \rho(x,t) \exp[i \phi(x,t)]$, decomposing the complex field $u(x,t)$ 
into its density $\rho(x,t)$ and phase $\phi(x,t)$; here, $u_{0}$ is an arbitrary complex constant. 
%
%
%
The resulting system of PDEs for $\rho(x,t)$ and  $\phi(x,t)$
possesses the exact (uniform) steady-state solution:
	\begin{equation}\label{6}
	\rho=1, \qquad \phi=2\sigma|u_{0}|^{2}t,
	\end{equation}
corresponding to the continuous-wave (cw): $u=u_{0}\exp(2i\sigma|u_{0}|^{2}t)$. 
The stability of the cw solution (studied also in Ref.~\cite{Arnaudon2}) can be investigated 
as follows. Let
	\begin{equation}\label{8}
	\rho=1+\tilde{\rho} , \qquad \phi=2\sigma|u_{0}|^{2}t+\tilde{\phi},
	\end{equation}  
where small perturbations $\tilde{\rho}$, $\tilde{\phi}$ are assumed to be $\propto \exp(ikx-i\omega t)$, 
while $k$ and $\omega$ denote the wavenumber and frequency. 
Substituting Eqs.~(\ref{8}) into the equations for $\rho(x,t)$ and  $\phi(x,t)$, we find that the latter 
obey the dispersion relation:
	\begin{equation}\label{9}
	\omega^{2}=\frac{k^{2}(-4\sigma|u_{0}|^{2}+k^{2})}{(1+a^{2}k^{2})^{2}}.
	\end{equation}
It is observed that in the case of the defocusing nonlinearity, i.e., for $\sigma=-1$, 
the cw solution is always modulationally stable, i.e., 
$\omega \in \mathbb{R}~\forall k\in \mathbb{R}$. On the other hand, for a 
focusing nonlinearity, $\sigma=+1$, the cw solution is unstable for $k^2<4|u_0|^2$: 
in this case, perturbations grow exponentially, with the 
instability growth rate given by Im$(k)$. Note that, for $a = 0$, Eq.~(\ref{9}) reduces 
to the well-known~\cite{kivshar,hasegawa} result for the modulational (in)stability 
of the NLS equation: $\omega^{2}=k^{2}(-4\sigma|u_{0}|^{2}+k^{2})$.	
Thus, as was also found in Ref.~\cite{Arnaudon2}, the interval of modulationally 
unstable wavenumbers is shared between NLS and CH-NLS and, in both cases, 
the defocusing realm of $\sigma=-1$ is modulationally stable.	

Below we consider the defocusing nonlinearity case and 
seek nonlinear excitations -- i.e., dark or antidark solitons -- which propagate 
on top of the (stable in this case) cw background. Notice that, for $\sigma=-1$, 
the long-wavelength limit ($k \rightarrow 0$) of Eq.~(\ref{9}) provides the so-called 
``speed of sound'', $\omega/k \equiv C=\pm 2u_{0}$ (with the sign $\pm$ denoting 
the propagation direction), namely the velocity of small-amplitude linear 
excitations propagating on top of the cw background.

\section{Asymptotic expansions and solitons}
	
\subsection{The Boussinesq equation}

We start our perturbation theory considerations by noting that the dispersion relation~(\ref{9}) 
is similar to the one of the Boussinesq equation, i.e., $\omega^{2} \sim k^{2}C^2+k^{4}$, 
(for $a^2 k^2 \ll 1$). This suggests the potential asymptotic reduction of 
the density and phase equations
to a Boussinesq-like equation.
Indeed, let us seek solutions 
in the form of the following asymptotic expansions:
	\begin{equation}\label{11}
	\phi=-2|u_{0}|^{2}t+\sqrt{\epsilon}\Phi, \qquad 
	\rho=1+\epsilon\rho_{1}+\epsilon^{2}\rho_{2}+ \cdots,
	\end{equation}
where $0 < \epsilon \ll 1$ is a formal small parameter, while the unknown real 
functions $\Phi$ and $\rho_{j}$ ($j=1,2,\ldots$) are assumed to depend on the 
slow variables $X$ and $T$ defined as:
	\begin{equation}\label{12}
	X=\sqrt{\epsilon}x, \qquad T=\sqrt{\epsilon}t.
	\end{equation}
        Substituting Eqs.~(\ref{11}) into
        the equations for $\rho(x,t)$ and  $\phi(x,t)$
        and using the variables in Eq.~(\ref{12}), 
we obtain the following results. First, the equation for the phase leads, at orders $\mathcal{O}(\epsilon^{3/2})$ and  $\mathcal{O}(\epsilon^{5/2})$, to the following equations:
	\begin{eqnarray}\label{13}
	\Phi_{T} +C^2 \rho_{1}=0,
	\end{eqnarray}
	\begin{eqnarray}\label{14}
	a^{2}\Phi_{XXT}-2|u_{0}|^{2}(2\rho_{2}+3\rho_{1}^{2}
	-a^{2}\Phi_{X}^{2})+\rho_{1XX}-\rho_{1}\Phi_{T}
	-\Phi_{X}^{2}=0.
	\end{eqnarray}
Second, the equation for the density leads at orders $\mathcal{O}(\epsilon)$ and $\mathcal{O}(\epsilon^{2})$ 
to the equations:
	\begin{eqnarray}\label{15}
	\Phi_{XX}+\rho_{1T}=0,
	\end{eqnarray}
	\begin{eqnarray}\label{16}
	\rho_{2T}+2a^{2}\Phi_{X}\Phi_{XT}+a^{2}C^{2}\Phi_{XX}\rho_{1}
	+a^{2}\Phi_{T}\Phi_{XX}-a^{2}\rho_{1XXT} 
	+2\Phi_{X}\rho_{1X}+\Phi_{XX}\rho_{1}=0.
	\end{eqnarray}

It is now possible to eliminate the functions $\rho_{1}$ and $\rho_2$ from the system 
of Eqs.~(\ref{13})-(\ref{16}), and derive the following equation for $\Phi(X,T)$:
	\begin{eqnarray}\label{17}
	\Phi_{TT}-C^{2}\Phi_{XX}+\epsilon \Big\{2a^{2}\Phi_{XXTT}
	 -4\Phi_{X}\Phi_{XT}(1-3a^{2}|u_{0}|^{2})
	 -2\Phi_{T}\Phi_{XX}
	-\Phi_{XXXX} 
	\Big \}= \mathcal{O}(\epsilon^{2}).
	\end{eqnarray}
At the leading-order, Eq.~(\ref{17}) is a second-order linear wave 
equation, while at order $\mathcal{O}(\epsilon)$ incorporates fourth-order
dispersion and quadratic nonlinear terms, resembling the 
Boussinesq and the Benney-Luke \cite{BL} equations. These models
have been used to describe bidirectional shallow water waves 
in the framework of small-amplitude and long-wavelength approximations;
see, e.g., the expositions
of~\cite{ablo2011,johnson74}
They were also used in other contexts including
ion-acoustic waves in plasmas \cite{infeld,karpman1},
mechanical lattices and electrical 
transmission lines \cite{rem}. 

It is worth mentioning that an analysis similar to that presented 
above can also be performed in two-dimensional (2D) settings: indeed, 2D Boussinesq equations 
were derived from 2D NLS equations with either a local \cite{pel1} or a nonlocal 
\cite{Hor2,Hor3} defocusing nonlinearity. Such studies are also
relevant to investigations 
concerning the
transverse instability of planar dark solitons \cite{pel2}.

\subsection{The KdV equation}

Next, using a multiscale expansion method similar 
to the one employed in the water wave problem \cite{ablo2011}, we will now
derive 
from the Boussinesq equation a pair of KdV equations 
for right- and left-going waves. These models will be obtained under the additional
assumption of unidirectional propagation. We thus seek solutions of 
Eq.~(\ref{17}) in the form of the asymptotic expansion:
\begin{equation}\label{19}
\Phi=\Phi_{0}+\epsilon\Phi_{1}+\cdots, 
\end{equation}
where the unknown functions $\Phi_j$ ($j=1,2,\ldots$) depend on the variables:
\begin{equation}\label{18}
\mathcal{\chi}=X-CT, \qquad \tilde{\mathcal{\chi}}=X+CT, \qquad \mathcal{T}=\epsilon T  
\end{equation}
(recall that $C^{2}=4|u_{0}|^{2}$). Substituting Eq.~(\ref{19}) into Eq.~(\ref{17}), 
and using Eq.~(\ref{18}), we obtain the following results. 
First, at the leading order, $\mathcal{O}(1)$, we obtain the wave equation:
	\begin{equation}\label{20}
	-4C^{2}\Phi_{0 \mathcal{\chi} \tilde{\mathcal{\chi}}}=0, 
	\end{equation}
which implies that $\Phi_{0}$ can be expressed as a superposition of a right-going wave, 
$\Phi_{0}^{ (R) }$, depending on $\mathcal{ \chi }$, and a left-going wave, $\Phi_{0}^{(L)}$, 
depending on $\tilde{ \mathcal{\chi} }$, namely:
	\begin{equation}\label{21}
	\Phi_{0}=\Phi_{0}^{(R)}+\Phi_{0}^{(L)}.
	\end{equation}
Second, at  $\mathcal{O}(\epsilon)$, we obtain:
	\begin{eqnarray}\label{22}
	4C^{2}\Phi_{1 \mathcal{\chi} \tilde{\mathcal{\chi}}} = C(3a^{2}C^{2}-2)(\Phi_{0 \mathcal{\chi}}^{(R)}\Phi_{0 \tilde{\mathcal{\chi}} \tilde{\mathcal{\chi}}}^{(L)}-
	\Phi_{0 \tilde{\mathcal{\chi}}}^{(L)}\Phi_{0 \mathcal{\chi} \mathcal{\chi}}^{(R)})
	\nonumber\\
	+ \Big[2C\Phi_{0 \mathcal{T}}^{(L)} -\frac{3C}{2}(2-a^{2}C^{2})\Phi_{0 \tilde{\mathcal{\chi}}}^{(L)2} -(1-2a^{2}C^{2})\Phi_{0 \tilde{\mathcal{\chi}} \tilde{\mathcal{\chi}} \tilde{\mathcal{\chi}}}^{(L)}  \Big]_{\tilde{\mathcal{\chi}}} 
	\nonumber\\
	-\Big[2C\Phi_{0 \mathcal{T}}^{(R)} -\frac{3C}{2}(2-a^{2}C^{2})\Phi_{0 \mathcal{\chi}}^{(R)2} +(1-2a^2 C^{2})\Phi_{0 \mathcal{\chi} \mathcal{\chi} \mathcal{\chi}}^{(R)}\Big]_{\mathcal{\chi}}. 
	\end{eqnarray}
It is now observed that when integrating Eq.~(\ref{22}) with respect to $\mathcal{\chi}$ or 
$\tilde{\mathcal{\chi}}$ secular terms arise from the square brackets in the right-hand 
side of this equation, which are functions of $\mathcal{\chi}$ or $\tilde{\mathcal{\chi}}$, not both. 
Hence, we set the secular terms to zero so as to avoid secular growth, and obtain two 
uncoupled nonlinear evolution equations for $\Phi_{0}^{(R)}$ and $\Phi_{0}^{(L)}$. 

Next, employing Eq.~(\ref{13}), it is straightforward to find that 
the amplitude $\rho_{1}$ can also be decomposed to a left- and a right-going wave, i.e.,  $\rho_{1}=\rho_{1}^{(R)}+\rho_{1}^{(L)}$, with
	\begin{equation}\label{23}
	\Phi_{0 \mathcal{\chi}}^{(R)}=C\rho_{1}^{(R)}, \quad
	\Phi_{0 \tilde{\mathcal{\chi}}}^{(L)}=-C\rho_{1}^{(L)}.
	\end{equation}
To this end, using the above equations for $\Phi_{0}^{(R)}$ and $\Phi_{0}^{(L)}$ yields 
the following equations for $\rho_{1}^{(R)}$ and $\rho_{1}^{(L)}$:
	\begin{eqnarray}\label{24}
	2C\rho_{1 \mathcal{T}}^{(R)} -3C^{2}(2-a^{2}C^{2})\rho_{1}^{(R)} \rho_{1 \mathcal{\chi}}^{(R)}+(1-2a^{2}C^{2})\rho_{1 \mathcal{\chi} \mathcal{\chi} \mathcal{\chi} }^{(R)} =0,
	\end{eqnarray}
	\begin{eqnarray}\label{25}
  2C\rho_{1 \mathcal{T}}^{(L)} +3C^{2}(2-a^{2}C^{2})\rho_{1}^{(L)} \rho_{1 \tilde{\mathcal{\chi}}}^{(L)} -(1-2a^{2}C^{2})\rho_{1 \tilde{\mathcal{\chi}} \tilde{\mathcal{\chi}} \tilde{\mathcal{\chi}} }^{(L)} =0.
	\end{eqnarray}
The above equations are two KdV equations for left- and right-going waves. Pertinent KdV soliton 
solutions will be used below for the construction of approximate soliton solutions of the CH-NLS equation.
	
	\subsection{Dark and antidark solitons of the CH-NLS equation}

To proceed further, we focus on the right-going wave and introduce the transformations: 
	\begin{eqnarray}\label{26}
	\mathcal{\hat{T}}=\Big(\frac{1-2 a^{2}C^{2}}{2C}\Big)\mathcal{T}, \qquad 
	\rho_{1}^{(R)}= \frac{2}{C^{2}}\left(\frac{1-2 a^{2}C^{2}}{2-a^{2}C^{2}}\right)U,
	\end{eqnarray}
to express the KdV Eq.~(\ref{24}) in its standard form:
	\begin{equation}\label{27}
	U_{\mathcal{\hat{T}}}-6UU_{\mathcal{\chi}}+U_{\mathcal{\chi} 
	\mathcal{\chi} \mathcal{\chi}}=0.
	\end{equation}
The above equation possesses the commonly known (see, e.g., Ref.~\cite{ablo2011}) soliton solution:
	\begin{equation}\label{28}
	U=-\frac{1}{2} \beta \; {\rm sech}^{2} \Big[\frac{\sqrt{\beta}}{2}(\mathcal{\chi}-\beta\mathcal{\hat{T}} +\mathcal{\chi}_{0}) \Big]
	\end{equation} 
where arbitrary constants $\beta>0$ and $\tilde{\mathcal{\chi}}_{0}$ set, respectively, the amplitude 
(as well as the width and velocity) and the initial position of the soliton. 
Using this solution and reverting transformations back to the original
variables,
we find the following approximate [valid up to order $\mathcal{O}(\epsilon)$] 
solution of Eq.~(\ref{1}):
	\begin{eqnarray}\label{29}
	u \approx u_{0}\Big[1- \frac{\epsilon \beta}{C^2} q~{\rm sech}^2(\xi) \Big] 
	\exp\Big[-2i|u_{0}|^{2} t 
	- i\frac{ \sqrt{\epsilon \beta} }{C}q \tanh(\xi) \Big], 
\end{eqnarray}
\begin{equation}
\xi=\frac{1}{2}\sqrt{\epsilon \beta} 
\left[ x-\left(C+\frac{\epsilon\beta(1-2a^{2} C^{2})}{2C} \right)t+x_{0} \right],
\label{31}
\end{equation}
where the parameter $q$ is given by:
	\begin{equation}\label{30}
	q=\frac{1-2a^{2}C^{2}}{2-a^{2}C^{2}}, 
	\end{equation}
and we have implicitly assumed that $C=2u_0$. The corresponding 
solution of Eq.~\eqref{25} (the left-moving soliton) is given by the same 
set of equations, but for $C=-2u_0$.
	 
At this point, it is important to mention that the approximate soliton solution~(\ref{30}) 
of the CH-NLS equation describes two types of solitons: for $q>0$ the solitons are dark 
(density dips on top of the cw background of amplitude $u_0$), while for $q<0$ the 
solitons are antidark (density humps on top of the cw background). The sign of parameter $q$ 
depends on the range of values of a single quantity
$p \equiv a^2 C^2 = 4a^2|u_0|^2$: 
indeed, if $1/2 < p <2$ the solitons are antidark, else they are dark. 
For a similar scenario, but in a different class of defocusing NLS 
models, see Refs.~\cite{kivshar90,yuriol,kiv91,kal,m1}, as well as Refs.~\cite{Hor1,Hor2,Hor3} 
for recent work on nonlocal media. 

\section{Numerical Exploration}

In this section we investigate the accuracy of our analytical predictions against 
direct numerical simulations performed by using 
the Exponential Time-Differencing 4th-order Runge-Kutta (ETDRK4) scheme 
of Refs.~\cite{7,8}. However, before proceeding further, it is relevant to discuss 
at first the specific methodology used.

\subsection{Numerical methods} 

First, we introduce the transformation $u(x,t)=u_{0}\exp\left(-2i|u_{0}|^{2} t \right)\psi(x,t) $, 
and rewrite the CH-NLS equation as:
\begin{equation}\label{32}
i \hat{m}_{t} + \psi_{xx} -2 |u_0|^2 \hat{m}(|\psi|^2-a^2 |\psi_{x}|^2 - 1) =0, 
\end{equation}
where $\hat{m} = \psi - a^2 \psi_{xx}$. As is also the case with the defocusing NLS, 
this has the advantage of removing the time-dependent phase factor from subsequent 
calculations (and hence also from the boundary conditions). It is then clear 
that the approximate soliton solution of Eq.~\eqref{32} [cf.~Eq.~(\ref{29})]:
\begin{equation}\label{33}
	\psi \approx \Big[1- \frac{\epsilon \beta}{C^2} q~{\rm sech}^2(\xi) \Big] 
	\exp\Big[-i\frac{ \sqrt{\epsilon \beta} }{C}q \tanh(\xi) \Big], 
\end{equation} 
representing an orbit homoclinic in the density with an
appropriate (heteroclinic orbit in the phase) profile.
While one can use inhomogeneous Dirichlet or homogeneous Neumann boundary
conditions to tackle this problem, here we follow a different path.
Due to our interest towards a spectral implementation of ETDRK4, we
seek to realize periodic boundary conditions. 
To reconcile that with the nature of our dark soliton solutions,
we solve the initial value problem (IVP) of Eq.~\eqref{32} 
with an initial condition $\psi(x,0)=\psi_0$ on the truncated spatial domain $[-L,L]$,
with the initial condition
\begin{equation}\label{37}
\psi(x,0)=\psi_0 \exp\left[-\left(\frac{x}{L^*}\right)^\gamma\right].
\end{equation}
Here, $L^*$, with $0 < L^* < L$, is the (sufficiently large) width of the background pulse, 
while the particular value of the exponent $\gamma \gg 1$ is not
especially important; here we use $\gamma=34$. 
Notice that for $x/L^* \ll  1$ the initial condition reduces to $\psi_0$ and thus 
the dynamics of the original IVP (on a smaller interval) can be obtained. Indeed, 
we have ensured that this is the case by considering different values of $L^*$ and $L$ 
and checking that the dynamics do not change. For the figures shown below,
we have used $L=2500$ and $L^*=1500$.
        
For all simulations below involving a single soliton (cf. Figs.~\ref{fig:Run1}-\ref{fig:Run3}), 
for $\psi_0$ we use:
\begin{eqnarray}\label{38}
\psi_0 = \left[1- \frac{\epsilon \beta}{C^2} q~{\rm sech}^2 (\xi_0)\right]
	\exp\left[-i\frac{ \sqrt{\epsilon \beta} }{C}q \tanh(\xi_0) \right], 
\end{eqnarray}
where $\xi_0 =\frac{1}{2}\sqrt{\epsilon \beta}(x + 100)$, 
$C=2u_0$ and $q$ as defined in Eq.~\eqref{30}.
On the other hand, for simulations involving two solitons (cf. Fig.~\ref{fig:Run4} 
except for panel~\ref{fig:Run4}(e) which will be explained below), 
we use for $\psi_0$ the product:
\begin{eqnarray}\label{40}
\psi_0 &=& 
\left[1- \frac{\epsilon \beta}{C_+^2} q_1 {\rm sech}^2 (\xi_+)\right]
	\exp\left[-i\frac{ \sqrt{\epsilon \beta} }{C_+}q_1 \tanh(\xi_+) \right]
\nonumber \\[.2cm] 
        &\times&
\left[1- \frac{\epsilon \beta}{C_-^2} q_2~{\rm sech}^2 (\xi_-)\right]
	\exp\left[-i\frac{ \sqrt{\epsilon \beta} }{C_-}q_2 \tanh(\xi_-) \right], 
\end{eqnarray}
where $\xi_{\pm} =\frac{1}{2}\sqrt{\epsilon \beta}(x \pm 200)$, $C_{\pm}=\pm 2u_0$, and  
$q_{1,2}=(1-2a_{1,2}^{2}C^{2})/(2-a_{1,2}^{2}C^{2})$.

At this point we should also note the following. In the single-soliton simulations, 
the same value of $a$ is used both for the PDE and the initial condition. On the other hand, 
for the two-soliton simulations, $a_1$ and $a_2$ appearing through $q_{1,2}$ in Eq.~(\ref{40}) 
are treated as independent parameters, not related to parameter $a$ involved in 
the CH-NLS equation. Thus, in the two-soliton collision case, we fix the value 
of $a$ in Eq.~(\ref{32}), as well as all other parameter values (along with $a_1$ and $a_2$) 
involved in the initial condition of Eq.~(\ref{40}), and integrate numerically 
Eq.~(\ref{32}) via the ETDRK4 integrator.
The rationale of this towards considering different
collision scenarios is discussed below.
                
\subsection{Results of direct simulations}

\begin{figure}[tbp]
\subfloat[]{\includegraphics[width=.45\textwidth]{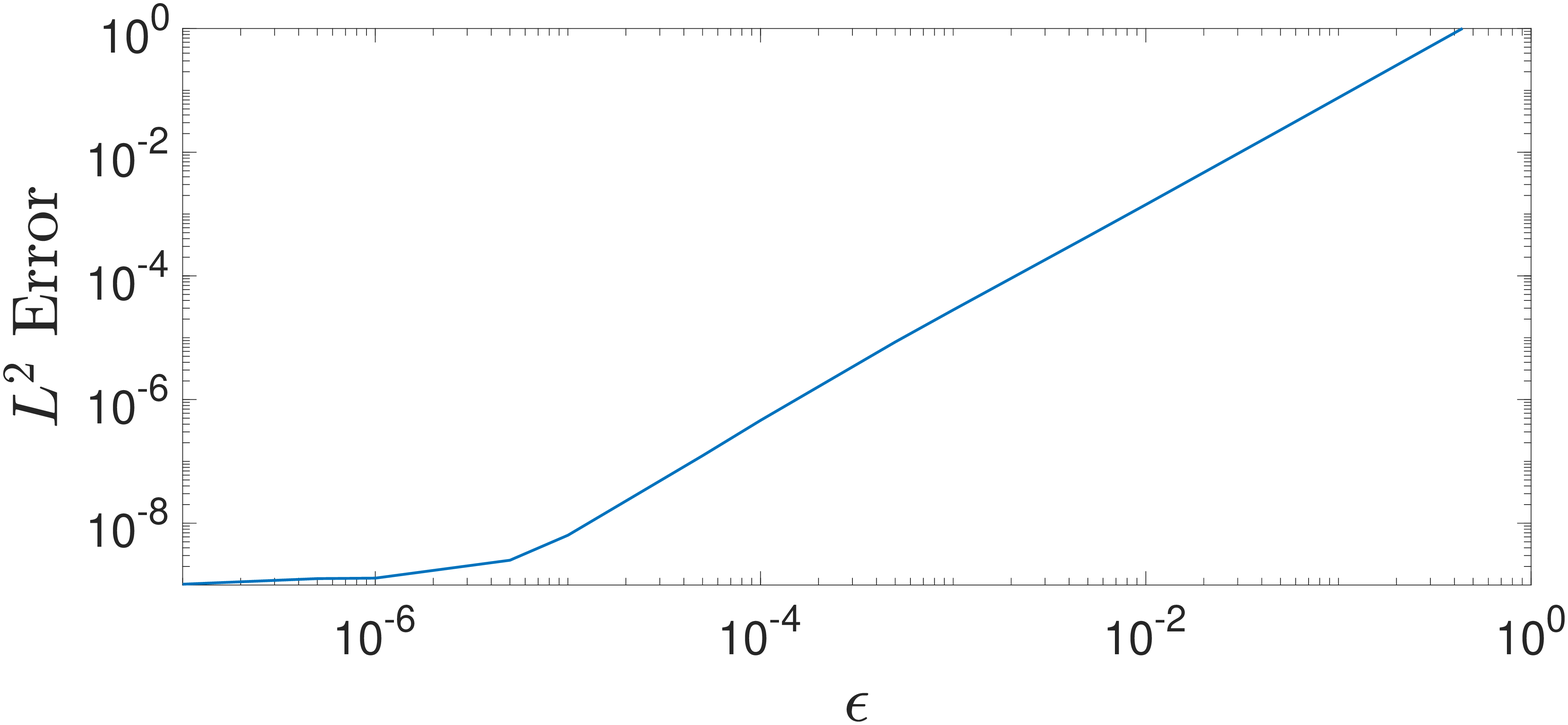}}

\subfloat[]{\includegraphics[width=.45\textwidth]{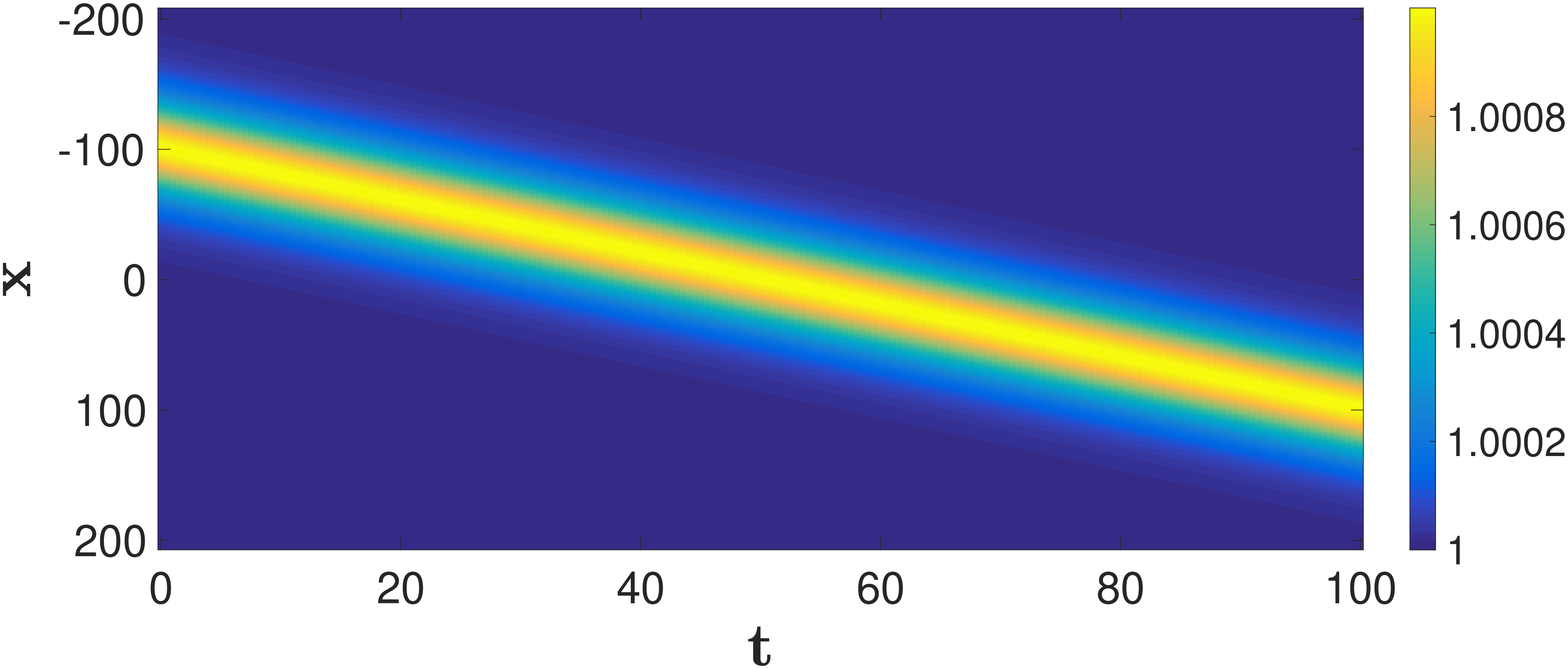}}
\subfloat[]{\includegraphics[width=.45\textwidth]{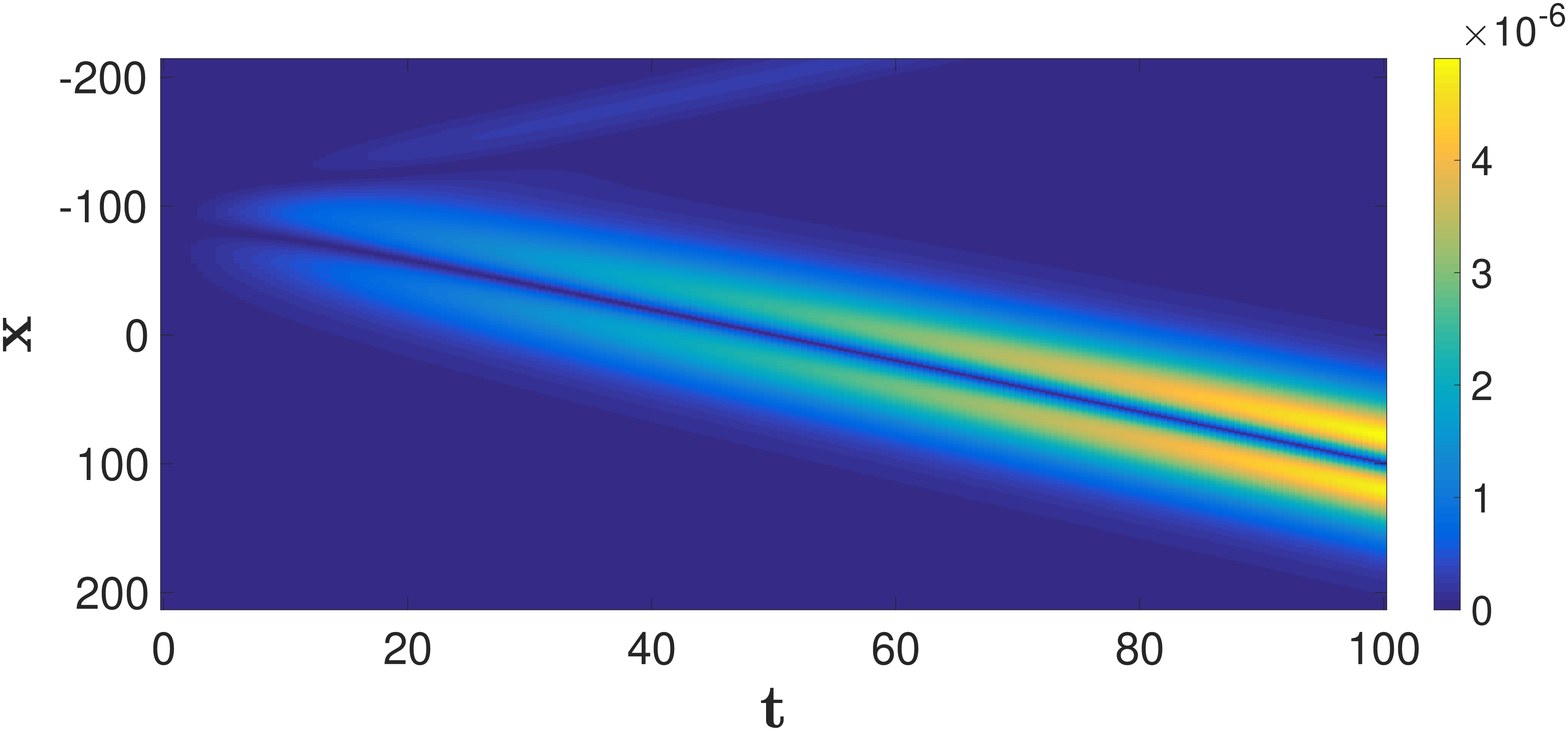}}

\caption{(a) Log-log plot of the $L^2$ error between the predicted solution \eqref{29} and the numerical solution as a function of $\epsilon$. All other parameter values fixed at $u_0=1$, $a=0.5$, $\beta=0.1$; 
in this case, parameter $p=1$, i.e., $p\in(1/2,2)$, which means that the soliton is antidark (cf. text).  
(b) Contour plot showing the evolution of the density of the corresponding antidark soliton, 
for $\epsilon =0.04$. (c) Same as (b) but for the difference between the densities 
of the analytical and the numerical solution. It can be observed that 
the error becomes larger as time increases; this is especially so at the ``wings'' (rather than 
the core) of the soliton. } 
\label{fig:Run1}
\end{figure}

First, our aim is to explore the validity of our asymptotic, small-amplitude soliton solution 
and its proximity to a true solution of the CH-NLS model. We thus fix all parameters 
in Eq.~\eqref{33} except for $\epsilon$, which we allow to vary. We then calculate the $L^2$ 
norm of the difference between the asymptotic solution and the numerical solution on 
the space-time domain $[-300,300] \times [0,100]$. 

\begin{figure}[htbp]

    \subfloat[]{\includegraphics[width=.45\textwidth]{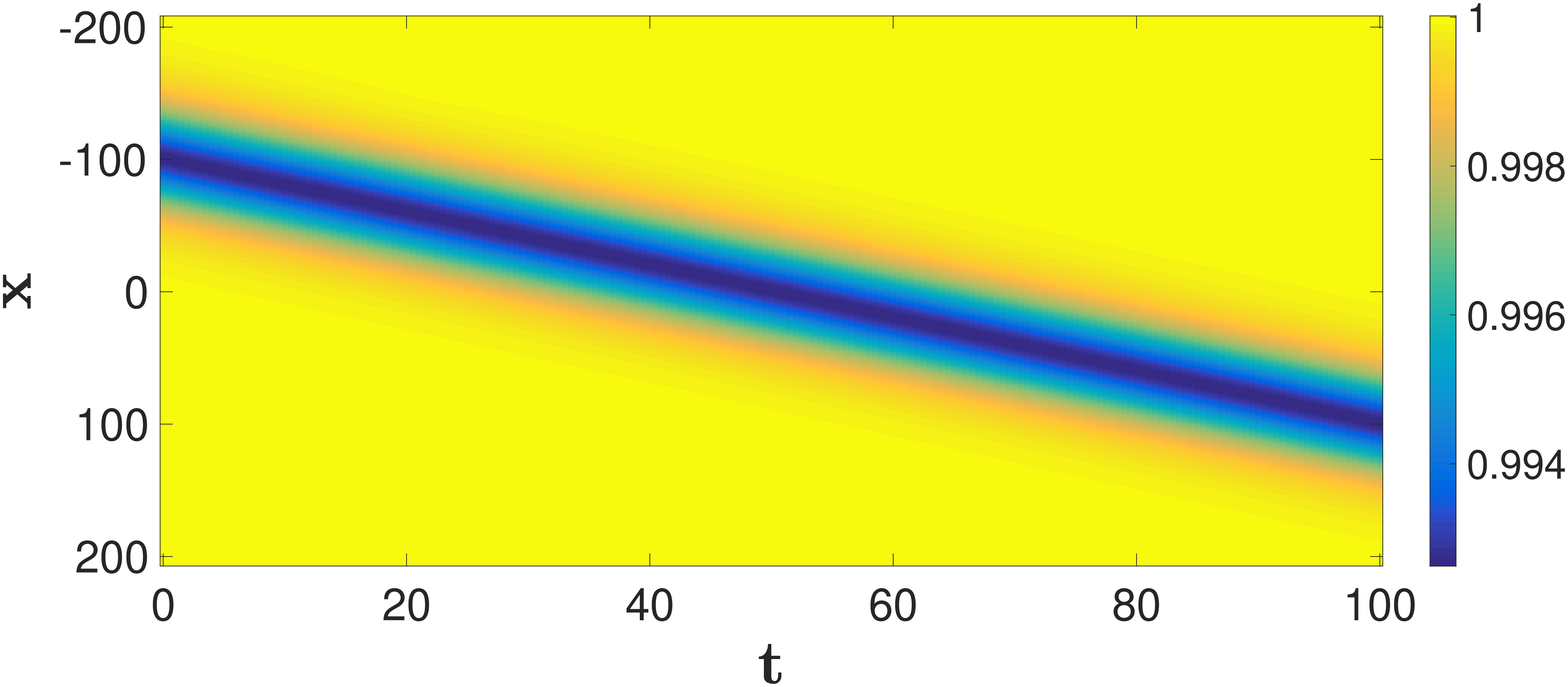}}
	\subfloat[]{\includegraphics[width=.45\textwidth]{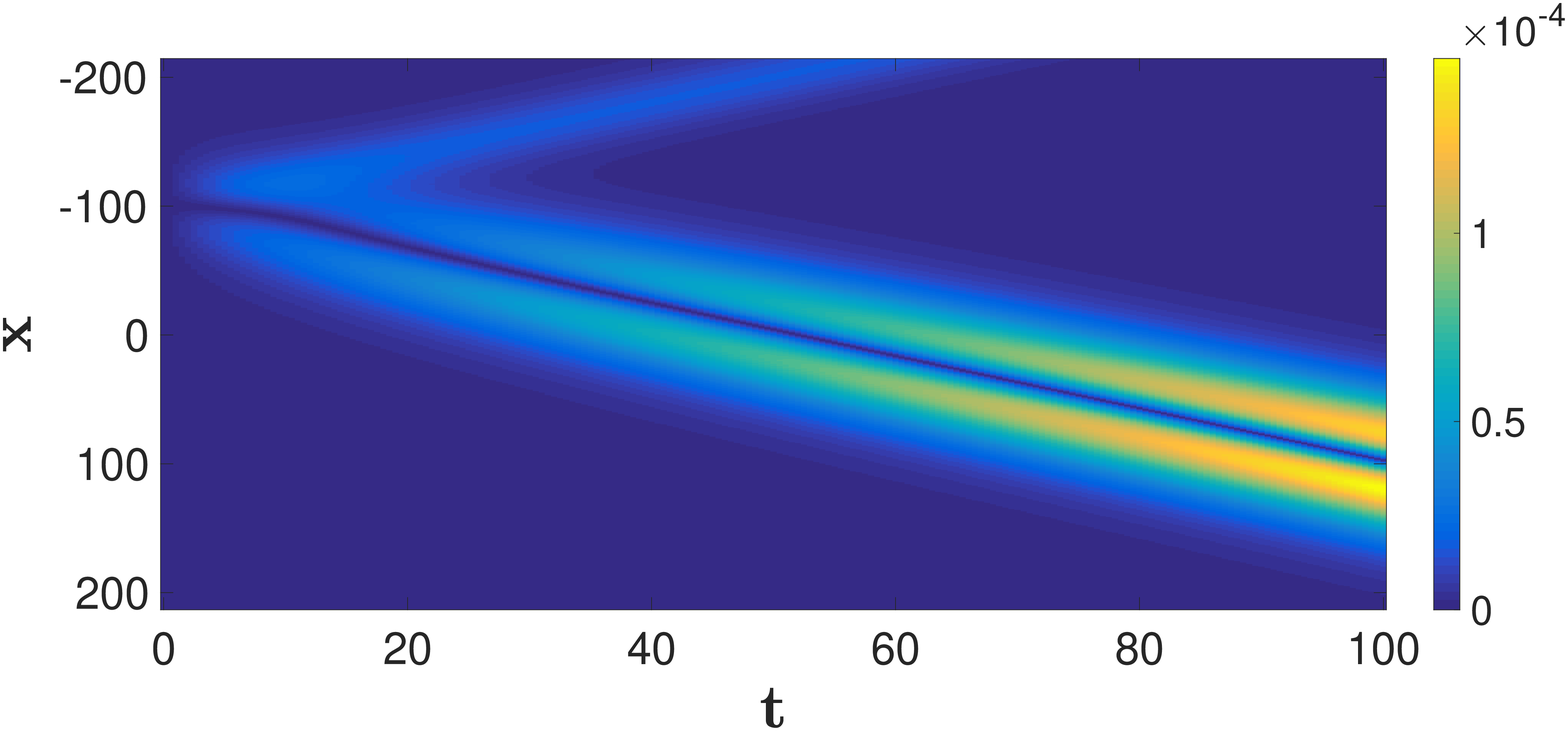}}

     \caption{(a) Same as Fig.~\ref{fig:Run1}(a), but for a dark soliton, for 
$u_0=1$, $a=0.8$, $\beta=0.1$, $\epsilon=0.04$; note that, here, $p=2.56>2$ and thus the 
soliton is indeed of the dark type (cf. text). (b) Same as Fig.~\ref{fig:Run1}(b), but for 
the dark soliton of panel (a). Again, the error becomes larger as time increases, 
especially at the ``wings'' of the soliton. } 
\label{fig:Run2}
\end{figure}

Figure~\ref{fig:Run1}(a) shows a log-log plot of the norm as a function of $\epsilon$; 
parameter values can be found in the caption, and correspond to an antidark soliton. 
As expected, the analytical and numerical solutions agree very well for sufficiently 
small $\epsilon$ (the error is at the order of numerical precision), but start to 
show differences once $\epsilon$ increases
appears to 
grow
in a way consistent with our theoretical expectations based on the expansion. 
The progressively increasing difference usually assumes the form of emerging 
linear radiative wavepackets. It may also have the more dramatic effect of splitting 
into a left-moving soliton and a right-moving soliton. Figure~\ref{fig:Run1}(b) 
shows a contour plot depicting the evolution of the density of the asymptotic 
antidark soliton solution, when $\epsilon$ is small. On the other hand, 
Fig.~\ref{fig:Run1}(c) shows a contour plot of the difference between the 
absolute value of the amplitudes of the numerical solution and the asymptotic solution.
Similar results have also been obtained for the case of dark solitons. In particular, 
Fig.~\ref{fig:Run2}(a) shows a contour plot depicting the evolution of the density of a 
dark soliton for $\epsilon=0.04$, while Fig.~\ref{fig:Run2}(b) compares the 
asymptotic and numerical solutions.
        
\begin{figure}[htbp]

    \subfloat[]{\includegraphics[width=.45\textwidth]{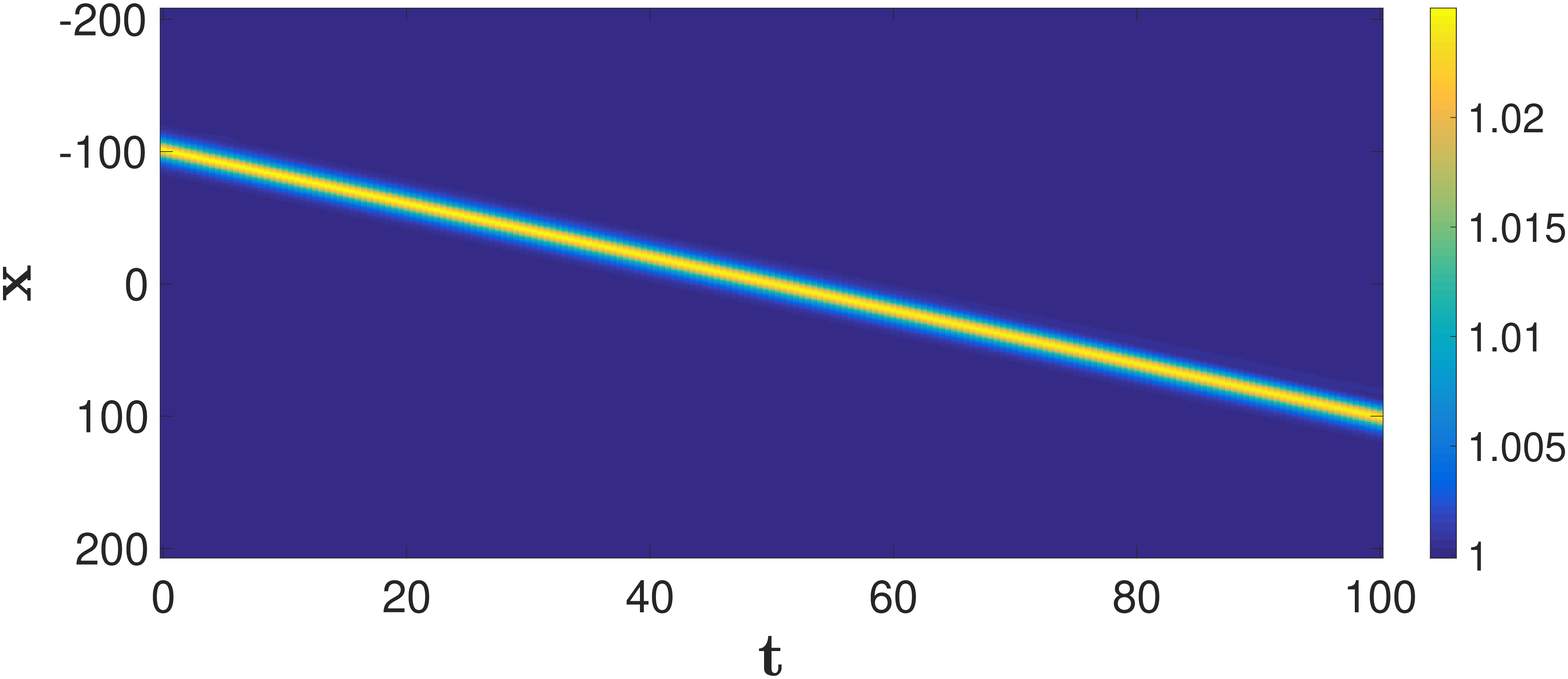}}
	\subfloat[]{\includegraphics[width=.45\textwidth]{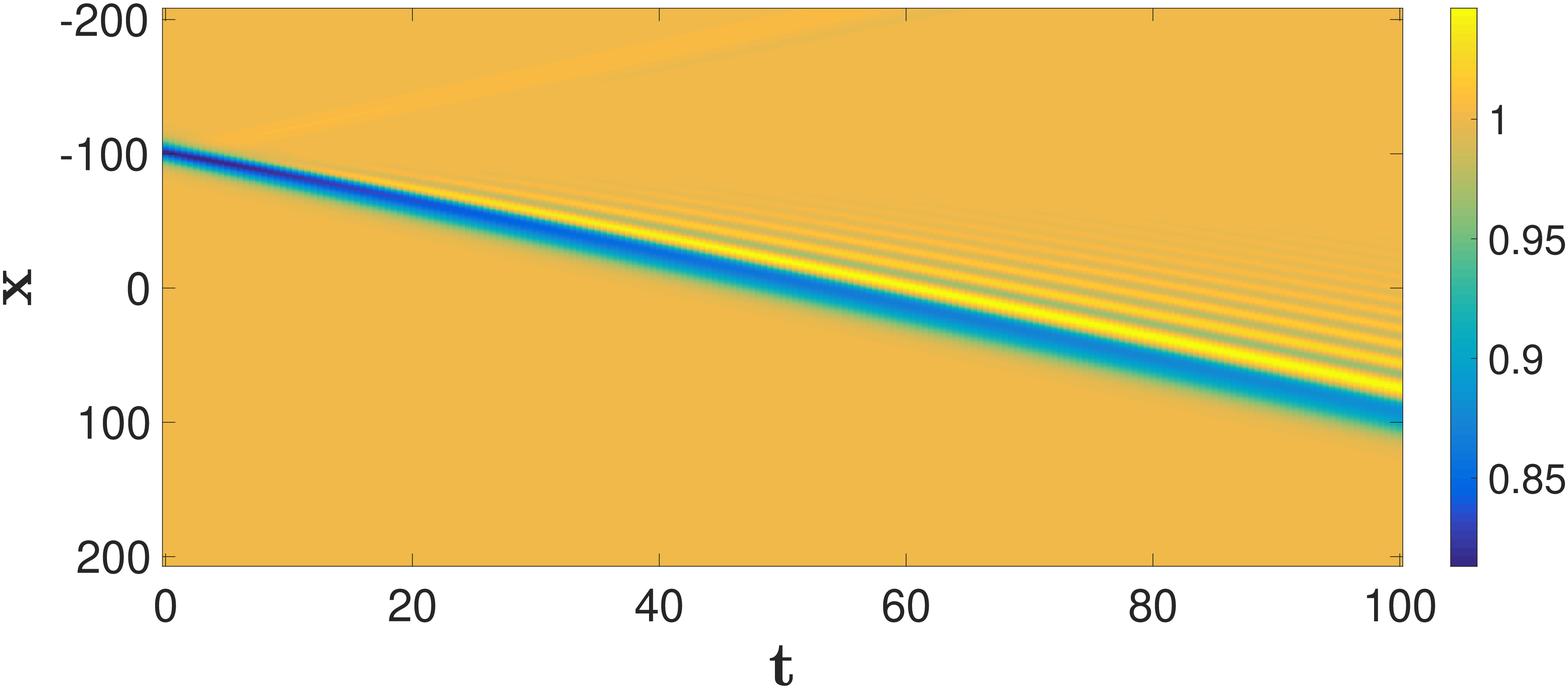}}

    \caption{Panels (a) and (b) are the corresponding large $\epsilon$ results of 
    Figs.~\ref{fig:Run1}(a) and \ref{fig:Run2}(a), respectively. Parameter values 
    are the same as in Figs.~\ref{fig:Run1}(a) and \ref{fig:Run2}(a) except that $\epsilon=1$.} 
     \label{fig:Run3}
\end{figure}

To test how the asymptotic analysis and the validity of the approximate soliton solutions fail, 
we have also examined the case where $\epsilon$ is quite large. Results of simulations for 
anti-dark and dark solitons are respectively illustrated 
in Figs.~\ref{fig:Run3}(a) and \ref{fig:Run3}(b), in the case of $\epsilon=1$; 
other parameter values are as in Figs.~\ref{fig:Run1} and \ref{fig:Run2}. 
It is observed that the antidark soliton of 
Fig.~\ref{fig:Run3}(a) agrees very well with the corresponding asymptotic solution 
(as the error in Fig.~\ref{fig:Run1}(a) suggests).
This clearly showcases the fact that the relevant parameter
controlling the amplitude of the solitary wave is
$\epsilon \beta q/C^2 =-0.025$ in this case, hence the
analytical approximation is still very accurate.
On the other hand, the dark soliton in Fig.~\ref{fig:Run3}(b) initially splits into 
a small-amplitude left-moving soliton (with no radiation) and into a right-moving one 
(with much radiation). We remark that Fig.~\ref{fig:Run3}(b) is fairly generic 
among the large $\epsilon$ solutions.

\begin{figure}[htbp]

	\subfloat[$a =a_1=a_2=0.75$, $b = 0.1$, $\epsilon=0.1$, $u_0 = 1$]{\includegraphics[width=.45\textwidth]{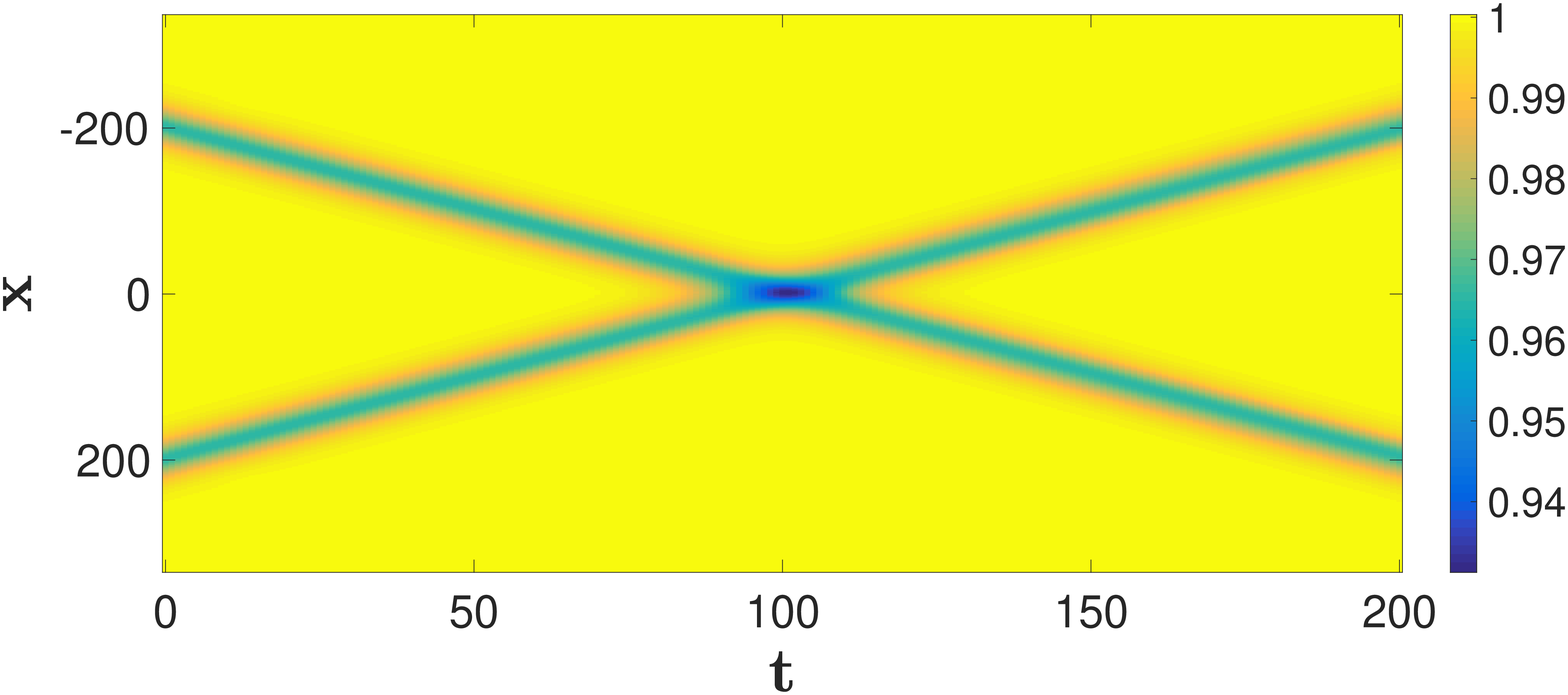}}
	\subfloat[$a =a_1=a_2=0.65$, $b = 0.1$, $\epsilon= 0.1$, $u_0 = 1$]{\includegraphics[width=.45\textwidth]{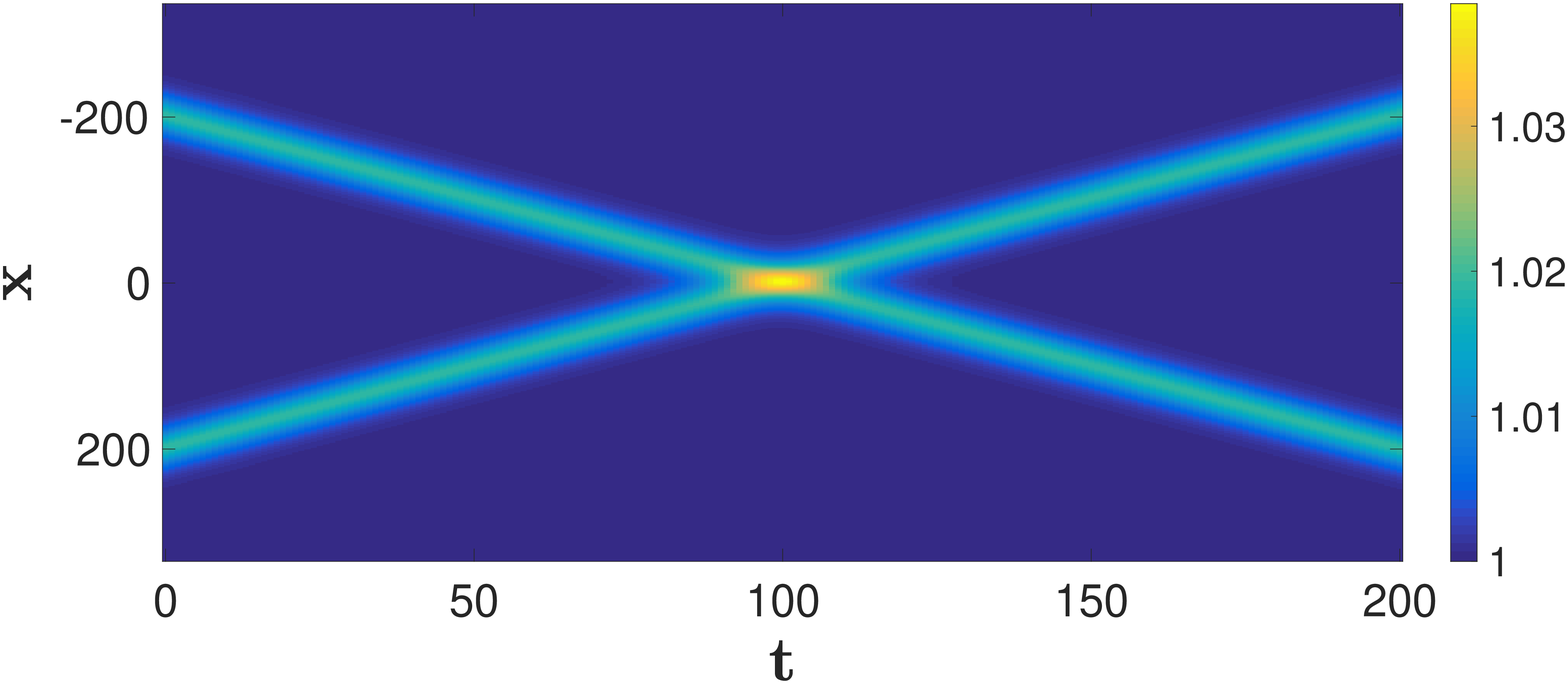}}
	
	\subfloat[$a =a_1=a_2=0.62$, $b = 0.1$, $\epsilon= 1$, $u_0 = 1$]{\includegraphics[width=.45\textwidth]{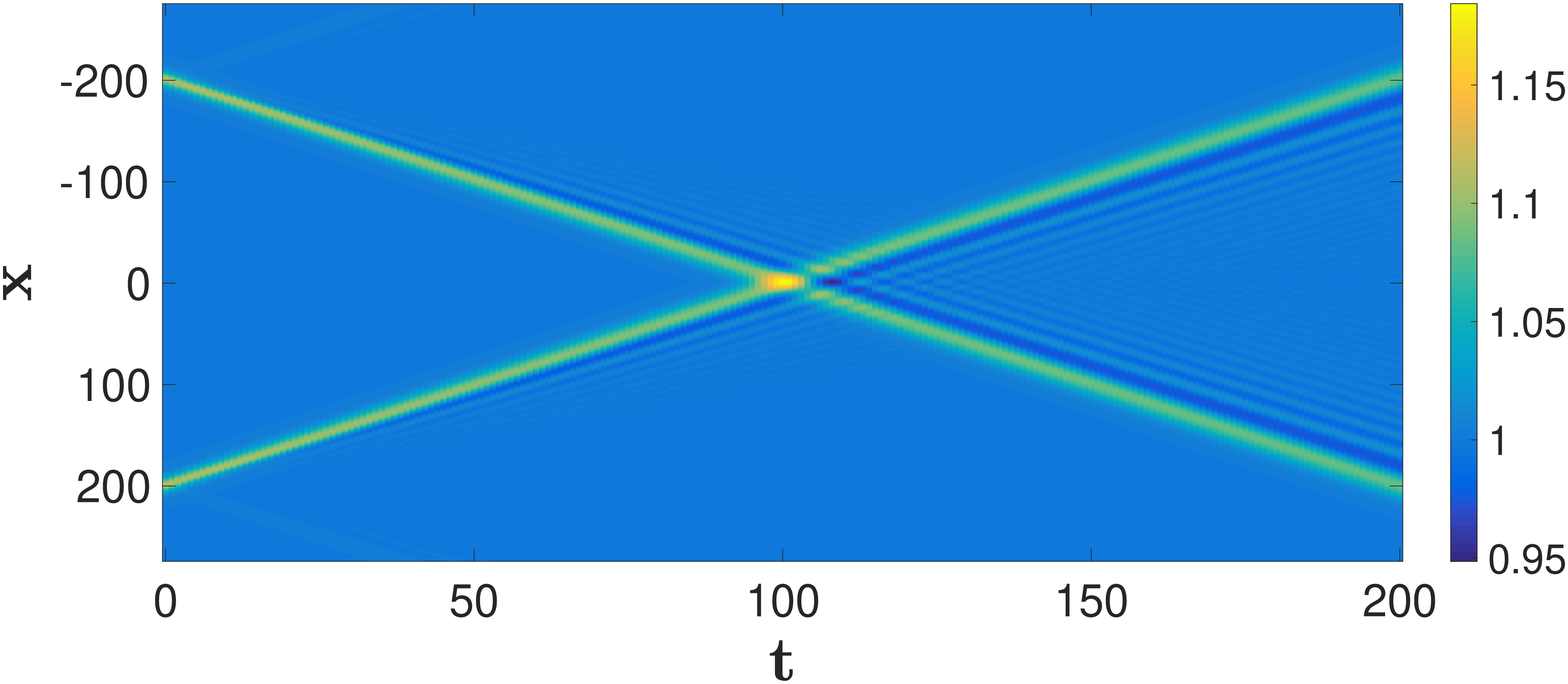}}
	
	\subfloat[$a=0.75$, $a_{1} = 0.67$, $a_{2} = 0.75$, $b = 0.1$, $\epsilon= 0.1$, $u_0 = 1$]{\includegraphics[width=.45\textwidth]{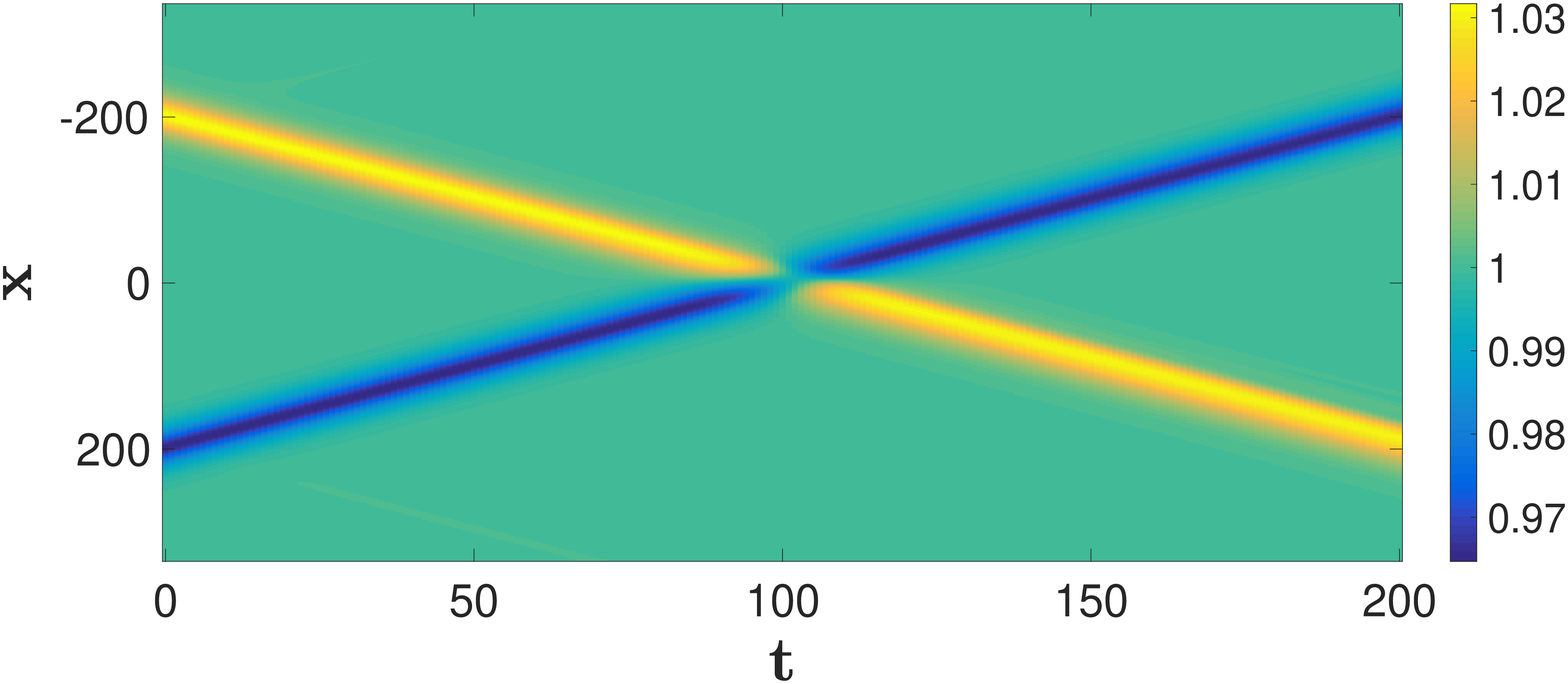}}
    \subfloat[$a=a_1=a_2=0.62$, $b = 0.1$, $\epsilon= 0.1$, $u_0 = 1$, $\nu=-0.7$]{\includegraphics[width=.45\textwidth]{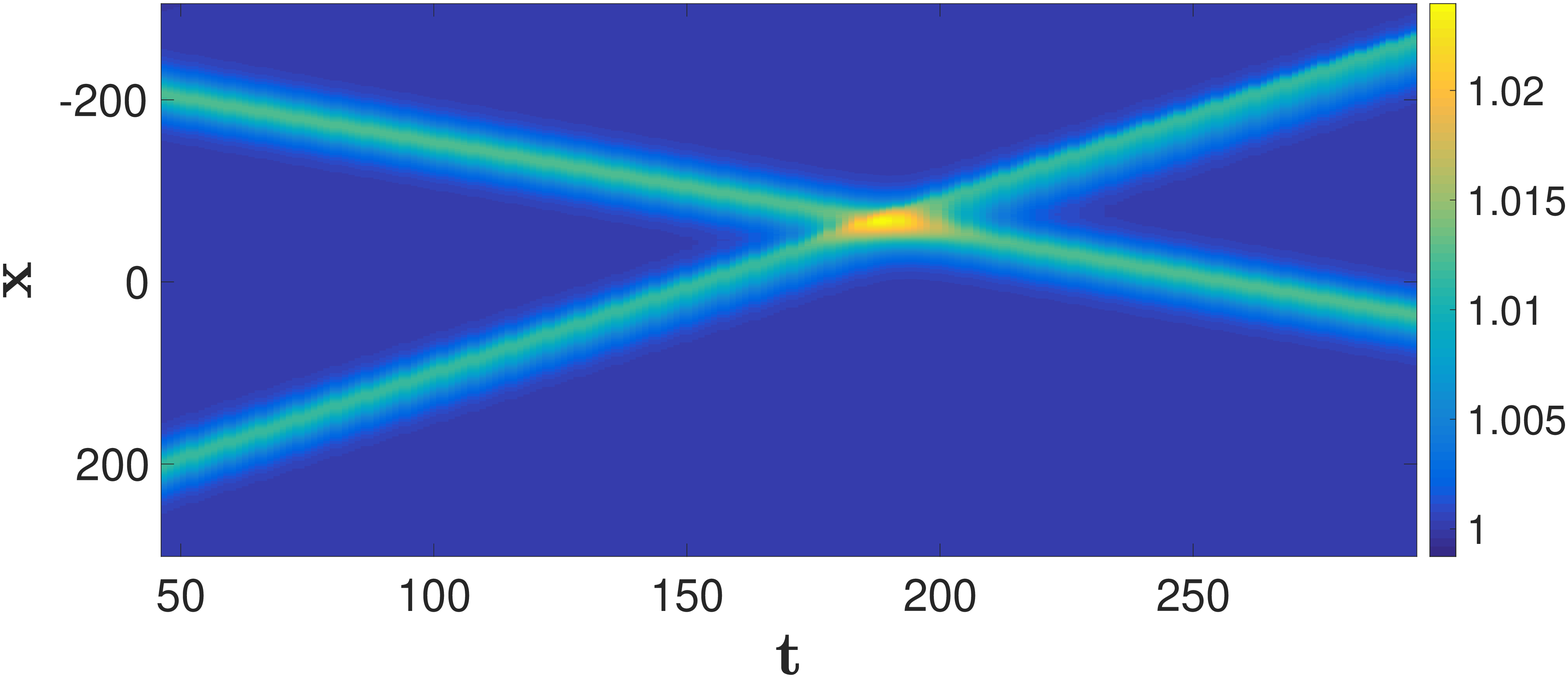}}

     \caption{Collisions of various dark and anti-dark solitons shown via contour plots of  there respective amplitudes. The intial condition used for each was Eq.~\eqref{40} with the given parameter values -- except (e) in which a gallilean boost was applied.} \label{fig:Run4}
\end{figure}

Next, in Fig.~\ref{fig:Run4}, we show numerical results for soliton collisions. 
In particular, Figs.~\ref{fig:Run4}(a) and \ref{fig:Run4}(b) depict, respectively, 
the collisions between two identical dark and two identical antidark solitons of 
opposite speeds, for $\epsilon=0.1$.  
As a contrast to Figs.~\ref{fig:Run4}(a) and \ref{fig:Run4}(b) where a relatively 
small value of $\epsilon$ was used, Fig.~\ref{fig:Run4}(c) shows the collision 
of two antidark solitons for $\epsilon=1$. It is worth noting that although emission of 
radiation is much stronger in this case, the solitons practically remain unscathed 
after their head-on collision. Here it should be mentioned that the initial 
condition for all the above three scenarios was Eq.~\eqref{40} with corresponding 
parameters found in the figure captions; particularly, in all cases, we used $a=a_1=a_2$.
        
As a final addition, we consider two special situations not predicted by 
the asymptotic analysis but found, rather, by purely numerical means.
Let us first recall [cf. Eq.~\eqref{30}] that antidark solitons exist when
$1/2 < p \equiv a^2 C^2 < 2$ and dark solitons otherwise. The asymptotic 
solution then implies that dark and antidark solitons should not coexist 
on the same range of $a$ values. Nevertheless, for values around the critical point 
$p=2$, one might expect that dark (or antidark) solitons may retain their 
form for relatively long times even if, technically speaking, they are not supported 
in the domain $p < 2$ (or $p > 2$). We have found that this is indeed the case: 
as shown in Fig.~\ref{fig:Run4}(d), a collision of an antidark soliton and 
a dark soliton, obtained for $a_1=0.67$, $a_2=0.75$ and $a=0.75$, is possible. 
It is observed that the antidark soliton appears to keep its form over 
the integration time (though it may very well decay over a
sufficiently long --considerably longer than the
horizon of the present dynamical considerations-- period).
        
The second situation deals with the fact that all previous solitons have had 
nearly the same speed. This is no accident as the soliton velocity $v$ is given by 
[cf. Eq.~\eqref{31}]:
        \begin{equation}\label{35}
        v=C+\frac{\epsilon\beta(1-2\alpha^{2} C^{2})}{2C}.
        \end{equation}
Notice that when $\epsilon$ and $\beta$ are small enough the velocity of the soliton 
approaches the speed of sound, i.e., $v \approx C=\pm 2|u_0|$. For instance, in the previous 
examples, we have used $u_0=1$ which makes the predicted soliton velocity $v \approx 2$
(as can also readily be observed in the figures). This fact implies that 
a collision between solitons of different speeds would not be possible because the
solitons live on top of the same background. 
%
%
Nevertheless, although the CH-NLS equation is not Galilean invariant, it is
conceivable (at least for small enough $a$) that a Galilean boost should increase 
the velocity of individual solitons. Indeed, Fig.~\ref{fig:Run4}(e) shows 
the asymmetric collision of two dark solitons obtained by using the
product of Eq.~\eqref{40} and $\exp(i \nu x)$ as $\psi_0$ [note that the initial soliton 
positions are $x_0=-250$ (left soliton) and $x_0=290$ (right soliton)]. 
Although not shown, the initial profile splits into a left-moving wave
and right-moving wave, except that the boost causes the 
left-moving wave to travel much faster, enabling a
collision with unequal speeds. We then observe that solitons of different 
speeds can coexist and that their collisions are also nearly elastic
in the small (solitary wave) amplitude regime considered.

\section{Conclusions \& Future Challenges}

Using asymptotic methods on the defocusing CH-NLS equation, 
we have derived an effective Boussinesq-type equation (to describe bi-directional
waves) and from there a pair of KdV models characterizing propagation 
in each of the one-dimensional directions. 
This has permitted us to systematically construct approximate
one-dimensional coherent structures in an explicit way of both
dark and anti-dark form, identifying their respective domains
of existence.

We have then used systematic numerical simulations to illustrate 
that these structures indeed persist in the full, original
CH-NLS model. Not only have we explored individual such structures,
but we have also considered pairs of them and observed them to emerge
practically unscathed as a result of their collisions in the
regime of (small) amplitudes considered.

These results offer a first glimpse into the possibilities of
the defocusing nonlinear realm for the case of the CH-NLS model.
Nevertheless, numerous open questions still remain in this
context. The technique used here only allows us to construct
small-amplitude structures. Yet, the question of whether 
``deep'' dark solitons (including black ones) exist and whether they 
are dynamically robust would require a different type of 
approach in order to be addressed. Additionally, whether
such structures may be identified in closed analytical form
is of interest in its own right. Furthermore, as far as we 
can tell, very little is known in either the focusing or 
the defocusing realm for the CH-NLS equation beyond the one-dimensional 
case. Questions including the potential collapse features, the 
existence and stability of vortices, among many others await 
further investigation and would be intriguing to explore in their 
own right. Relevant topics are currently under examination and 
will be reported in future studies.
\\ \\
P.G.K. gratefully acknowledges support from NSF-PHY-1602994, as well as  from FP7, Marie Curie Actions, People, International Research Staff Exchange Scheme (IRSES-605096); 
I. K. M and V. R. gratefully acknowledges support from FP7, from FP7, Marie Curie Actions, People, International Research Staff Exchange Scheme (IRSES-605096);

\end{document}